\documentclass[prl,amsmath,a4paper,twocolumn]{revtex4}
\usepackage{graphics}
\usepackage{amsfonts}

\newcommand{\ket}[1]{\left|#1\right\rangle}
\newcommand{\bra}[1]{\left\langle#1\right|}
\newcommand{\tr}{{\rm Tr}}

\begin{document}

\title{Nonadditive quantum error-correcting code}
\author{Sixia Yu$^{1,2}$, Qing Chen$^1$, C. H. Lai$^2$ and C. H. Oh$^2$}
\affiliation{$^1$Hefei National Laboratory for Physical Sciences
at Microscale and Department of Modern Physics \& Department of
Modern Physics University
of Science and Technology of China, Hefei 230026, P.R. China\\
$^2$Department of Physics, National University of Singapore, 10 Kent
Ridge Crescent,  Singapore 119260}
\begin{abstract}
We report the first nonadditive quantum error-correcting code,
namely, a $((9,12,3))$ code which is a 12-dimensional subspace
within a 9-qubit Hilbert space, that outperforms the optimal
stabilizer code of the same length by encoding more levels while
correcting arbitrary single-qubit errors.
\end{abstract}
\pacs{03.67.Pp} \maketitle

The quantum error-correcting code (QECC)
\cite{shor,ben,steane,knill} provides an active way of protecting
our quantum data from decohering. Almost all the QECCs constructed
so far are stabilizer codes \cite{gottesman, cal1,cal2}, codes that
have the structure of an eigenspace of an Abelian group generated by
mulitilocal Pauli operators. Codes without such a structure are
called nonadditive codes. The first nonadditive code
\cite{rains1,rains2} that outperforms the stabilizer codes is the
$((5,6,2))$ code, a 5-qubit code encoding 6 levels capable of
correcting single-qubit {\em erasure}, i.e., a code of distance 2.
Recently a family of distance 2 nonadditive codes with a higher
encoding rate has been constructed \cite{smolin}. Though some
nonadditive error-correcting codes had been constructed
\cite{nonadd,nonadd2}, the question of whether the nonadditive
error-correcting codes with a distance larger than 2 can encode more
levels than the corresponding stabilizer codes remains open.

In this Letter we report the first nonadditive code of distance 3
that beats the corresponding stabilizer code: a nonadditive
$((9,12,3))$ code that is a 12-dimensional subspace in a 9-qubit
Hilbert space against arbitrary single-qubit errors. In comparison,
the best stabilizer code $[[9,3,3]]$ of the same length can encode
only 3 logical qubits, i.e., an 8-dimensional subspace \cite{cal2}.

Our new code is most conveniently formulated in terms of
graph states \cite{werner,graph}. Let $G=(V,\Gamma)$ be an undirected simple graph  with
$|V|=n$ vertices and $\Gamma$, called as the {\it adjacency matrix} of the graph,
is an $n\times n$ symmetric matrix with vanishing diagonal entries
and $\Gamma_{ab}=1$ if vertices $a,b$ are connected and
$\Gamma_{ab}=0$ otherwise. Consider a system of $n$ qubits labeled
by $V$ and denote by $\mathcal X_a,\mathcal Y_a$, and $\mathcal Z_a$ three
Pauli operators acting on qubit $a\in V$. The {\it graph state} associated with graph $G$ reads
\begin{equation}\label{g}
\ket
G=\prod_{\Gamma_{ab}=1}\mathcal U_{ab}\ket+^V_x=\frac1{\sqrt{2^n}}\sum_{\vec\mu=\bf
0}^{\bf 1}
 (-1)^{\frac12\vec\mu\cdot\Gamma\cdot\vec\mu}\ket{\vec\mu}_z,
\end{equation}
where $\ket{\vec\mu}_z$ is the common eigenstates of $\{\mathcal
Z_a\}_{a\in V}$ with $(-1)^{\mu_a}$ as eigenvalues, $\ket +_x^V$
denotes the simultaneous +1 eigenstate of $\{\mathcal X_a\}_{a\in
V}$, and $\mathcal U_{ab}=(1+\mathcal Z_a+\mathcal Z_b-\mathcal
Z_a\mathcal Z_b)/2$ is the controlled-phase operation between qubit
$a$ and $b$. The graph state is also the unique simultaneous +1
eigenstate of $n$ vertex stabilizers $\mathcal G_a=\mathcal
X_a\mathcal Z_{N_a}$ with $a\in V$ where $N_a$ is the neighborhood
of $a$ and we denote by $\mathcal Z_{U}=\prod_{a\in U}\mathcal Z_a$
for a subset of vertices $U\subseteq V$.

\begin{figure}
\includegraphics{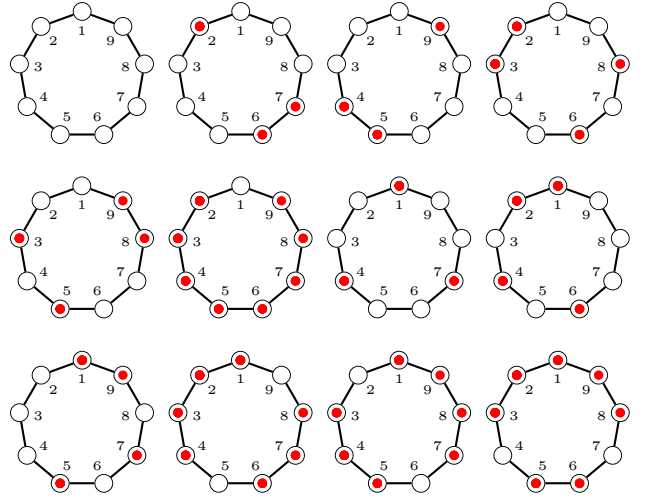}
\caption{(Color online) Twelve graph-state bases on the loop graph
$L_9$ for the $((9,12,3))$ code $\mathbb D$. Each graph represents a
graph state that is the unique common eigenstate of the vertex
stabilizers $\{\mathcal G_a\}$ with eigenvalue +1 if $a$ is
uncolored and -1 if the vertex is red-colored.}
\end{figure}
We consider in what follows the loop graph $L_9$ on 9 vertices which
are labeled by integers from 1 to 9. Its adjacency matrix has
nonvanishing entries $\Gamma_{aa_\pm}=1$ $(1\le a\le 9)$ only where
$a_\pm=a\pm1$  with identifications $9_+=1$ and $1_-=9$. The
corresponding graph state is denoted as $\ket{L_9}$. We claim that
the 12-dimensional subspace $\mathbb D$ spanned by the states
 $\{\mathcal Z_{V_i}\ket{L_9}\}_{i=1}^{12}$ where
\begin{eqnarray}
&V_1=\emptyset,\; V_2=\{2,6,7\},\; V_3=\{4,5,9\},\;
V_4=\{2,3,6,8\}\cr &V_5=\{3,5,8,9\},\; V_6=\{2,3,4,5,6,7,8,9\}&\cr
&V_7=\{1,4,7\},\;V_8=\{1,2,4,6\},\;V_9=\{1,5,7,9\}&\cr
&V_{10}=\{1,2,3,4,6,7,8\},\;V_{11}=\{1,3,4,5,7,8,9\}&\cr
&V_{12}=\{1,2,3,5,6,8,9\},&\end{eqnarray} as shown in Fig.1, is a
$((9,12,3))$ code. Obviously these 12 states are mutually orthogonal
since $V_i's$ are distinct and $\bra{G}\mathcal
Z_V\ket{G}=\delta_{V,\emptyset}$ holds true for any graph state. To
prove that the code is of distance 3, i.e., capable of correcting
single-qubit errors, we have only to demonstrate that each one of
$3\times 9$ single-qubit errors and $9\times 36$ two-qubit errors
$\mathcal E$ will bring $\mathbb D$ into its orthogonal complement
\cite{knill,werner}, i.e.,
\begin{equation}\label{cond}
\bra{L_9}\mathcal Z_{V_i}\mathcal E\mathcal Z_{V_j}\ket{L_9}=0, \quad (1\le i,j\le 12).
\end{equation}

Since all the bases of $\mathbb D$ given above are the common
eigenstates of the vertex stabilizers $\{\mathcal G_a=\mathcal
Z_{a_-}\mathcal X_a\mathcal Z_{a+}\}_{a=1}^9$, a bit flip error
$\mathcal X_a$ on these bases is equivalent to a phase flip error
$\mathcal Z_{N_a}$ on qubits in its neighborhood, e.g.,
$N_a=\{a_+,a_-\}$ in $L_9$, upto an unimportant phase factor. And a
$\mathcal Y_a$ error can be equivalently replaced by a phase flip
error $\mathcal Z_a\mathcal Z_{N_a}$ on quibts $a, a_+,$ and $a_-$.
As a result every single-qubit error is equivalent to one of the
following phase flip errors $\{\mathcal Z_a,\mathcal
Z_{N_a},\mathcal Z_a\mathcal Z_{N_a}\}$ for $1\le a\le 9$ and every
two-qubit error is equivalent to one of the following phase flip
errors
\begin{equation}
\begin{matrix}
\mathcal Z_a\mathcal Z_b, & \mathcal Z_{N_a}\mathcal Z_b, & \mathcal Z_{N_a}\mathcal Z_a\mathcal Z_b,\cr
\mathcal Z_a\mathcal Z_{N_b}, & \mathcal Z_{N_a}\mathcal Z_{N_b},&\mathcal Z_a\mathcal Z_{N_a}\mathcal Z_{N_b},\cr
\mathcal Z_a\mathcal Z_b \mathcal Z_{N_b},& \mathcal Z_{N_a}\mathcal Z_{N_b}\mathcal Z_b, &
\mathcal Z_a\mathcal Z_b\mathcal Z_{N_a}\mathcal Z_{N_b},\cr
\end{matrix}
\end{equation}
with $1\le a,b\le 9$. To summarize, for a loop graph, every
single-qubit or two-qubit error is equivalent to  one of following 6
patterns of phase flip errors
\begin{eqnarray}\label{type}
{\rm I}: && \mathcal Z_a, \cr
{\rm II}:&& \mathcal Z_a\mathcal Z_b,\cr
{\rm III}:&& \mathcal Z_{a_-}\mathcal Z_b\mathcal Z_{a_+},\; \mathcal Z_{a_\pm}\mathcal Z_{a}\mathcal Z_{a\pm3},\cr
{\rm IV}: && \mathcal Z_{a_-}\mathcal Z_{a_+}\mathcal
Z_{b_-}\mathcal Z_{b_+},\;
\mathcal Z_{a_-}\mathcal Z_{a}\mathcal Z_{a_+}\mathcal Z_{b},\cr && \mathcal Z_{a_-}\mathcal Z_{a-2}\mathcal Z_{a_+}\mathcal
Z_{a+2},\cr
{\rm V}: && \mathcal Z_{a_-}\mathcal Z_{a}\mathcal
Z_{a_+}\mathcal Z_{b_-}\mathcal Z_{b_+},\cr
{\rm VI}:&& \mathcal Z_{a_-}\mathcal Z_{a}\mathcal Z_{a_+}\mathcal Z_{b_-}\mathcal
Z_{b}\mathcal Z_{b_+},
\end{eqnarray}
where $a,b$ are suitably chosen so that error patterns I, II,
III, IV, V, VI are phase flip errors on 1 qubit to 6
qubits respectively. It is clear that phase flip errors on more than
6 qubits cannot be caused by any single-qubit or two-qubit error.

As an immediate consequence,  Eq.(\ref{cond}) is equivalent to
saying that {\em none} of the transition operators $\mathcal
Z_{V_i}\mathcal Z_{V_j}$ ($1\le i< j\le 12$) between each pair of
bases of $\mathbb D$ belongs to any one of 6 error patterns listed
in Eq.(\ref{type}). Because  $\mathcal Z_{V_k}\mathcal
Z_{V_7}=\mathcal Z_{V_{k+6}}$ it is enough to examine the following
31 different transition operators
\begin{equation*}
\begin{tabular}{lllllll}
$\mathcal Z_{147},$&$\mathcal Z_{126},$& $\mathcal Z_{1246},$ & $\mathcal Z_{2368},$& $\mathcal Z_{12569},$   &$\mathcal Z_{1234678},$ &$\mathcal Z_{12345689},$  \cr
                  &$\mathcal Z_{159},$& $\mathcal Z_{1348},$ & $\mathcal Z_{2569},$& $\mathcal Z_{23678},$   &$\mathcal Z_{1235689},$ &$\mathcal Z_{12356789},$  \cr
                  &$\mathcal Z_{267},$& $\mathcal Z_{1378},$ & $\mathcal Z_{3589},$& $\mathcal Z_{34589},$   &$\mathcal Z_{1245679},$ &$\mathcal Z_{23456789},$  \cr
                  &$\mathcal Z_{348},$& $\mathcal Z_{1579},$ &                    & $\mathcal Z_{123468},$  &$\mathcal Z_{1345789},$ &  \cr
                  &$\mathcal Z_{378},$& $\mathcal Z_{2467},$ &                    & $\mathcal Z_{135789},$  &$\mathcal Z_{2345689},$ &  \cr
                  &$\mathcal Z_{459},$& $\mathcal Z_{4579},$ &                    & $\mathcal Z_{245679},$  &$\mathcal Z_{2356789}.$  & \cr
\end{tabular}
\end{equation*}
obtained from $\mathcal Z_{V_7}$ and $\{\mathcal Z_{V_i}\mathcal
Z_{V_j},\mathcal Z_{V_7} \mathcal Z_{V_i}\mathcal Z_{V_j}|1\le i<
j\le 6\}$. It is easy to check that phase flip errors on 5 or more
qubits in the above table do not belong to any one of the error
patterns in Eq.(\ref{type}). Because of the symmetry of the loop
graph $L_9$, one needs only to check that $\mathcal Z_{126}$,
$\mathcal Z_{147}$,$\mathcal Z_{1246}$, and $\mathcal Z_{2368}$ do
not belong to any one of the error patterns in Eq.(\ref{type}),
which are easy tasks to perform. In this way we have demonstrated
that $\mathbb D$ is a ((9,12,3)) code, which is obviously
nonadditive.

As to  the projector of the code $\mathbb D$, we notice that there
are  3 local stabilizers of the code $\mathbb D$, namely, $\mathcal
G_{38},\mathcal G_{62}$, and $\mathcal G_{95}$, where we have
denoted $\mathcal G_{U}=\prod_{v\in U}\mathcal G_v$ for a subset of
vertices $U$. By denoting
\begin{eqnarray}
\mathcal A&=&\mathcal G_{14}
\Big(1-\mathcal G_{36}+\mathcal G_{39}-\mathcal G_{69}+
2\mathcal G_{369}+2\mathcal G_9\Big)\cr
&&+\mathcal G_{17}\Big(1-\mathcal G_{39}+\mathcal G_{36}-\mathcal G_{69}+
2\mathcal G_{369}+2\mathcal G_6\Big),
\end{eqnarray}
we can write down the projector of the code $\mathbb D$ as
\begin{equation}
\mathcal P=\frac1{2^{10}}(1+\mathcal G_{38})(1+\mathcal
G_{62})(1+\mathcal G_{95})\mathcal A(\mathcal A+8),
\end{equation}
from which the weight enumerator \cite{enum,rains3} of the code
$\mathbb D$ can be readily obtained
\begin{equation}
2^9\times
12\times\left(\frac{3u^9}{128}+\frac{u^5v^4}{64}+\frac{u^3v^6}{4}+\frac{u^2v^7}2+\frac{27uv^8}{128}\right).
\end{equation}
Here the coefficients of $u^{9-d}v^d$ is given by $\sum\tr^2
(\mathcal P \mathcal E_d)$ with the summation
 taken over all
(Hermitian) errors acting nontrivially on $d$ qubits.

To conclude, we have provided the first evidence that nonadditive
error-correcting codes can perform better than the stabilizer codes.
Since the bases of our code are all graph states, they can be easily
be prepared from a product state by using controlled phase operation
and local unitary operations as shown in Eq.(\ref{g}).

Y.S. acknowledges the financial support of NNSF of China (Grant No.
90303023 and Grant No. 10675107) and the ASTAR grant
R-144-000-189-305.

\end{document}